\documentclass{emulateapj}
\usepackage{txfonts}
\usepackage{color}
\slugcomment{The Astrophysical Journal, 705:L58, 2009}
\shorttitle{Helium Sedimentation and the UV Upturn}
\shortauthors{Peng \& Nagai}

\newcommand{\unitspace}{\ensuremath{\,}}
\newcommand{\usp}{\unitspace}

\newcommand{\mb}{\ensuremath{m_\mathrm{p}}} 

\newcommand{\beq}{\begin{equation}}
\newcommand{\eeq}{\end{equation}}
\newcommand{\beqa}{\begin{eqnarray}}
\newcommand{\eeqa}{\end{eqnarray}}

\newcommand{\gas}{_{\mathrm{gas}}}

\newcommand{\he}{_{\mathrm{He}}}
\newcommand{\hy}{_{\mathrm{p}}}

\def\r500{\ensuremath{r_{\text{500}}}}
\def\M500{\ensuremath{M_{\text{500}}}}
\def\T500{\ensuremath{T_{\text{500}}}}

\begin{document}

\title{Helium Sedimentation and the Ultraviolet Upturn in Brightest Cluster Galaxies}
\author{Fang Peng\altaffilmark{1}, Daisuke Nagai\altaffilmark{2}}
\altaffiltext{1}{Theoretical Astrophysics, 1200 E California Blvd, 350-17, Pasadena, CA 91125; fpeng@caltech.edu}
\altaffiltext{2}{Department of Physics,Yale University, P.O. Box 208210, New Haven, CT 06520-8120; daisuke.nagai@yale.edu}

\begin{abstract}
Recent observations with Galaxy Evolution Explorer (GALEX) show strong unexpected UV excess in the spectrum of brightest cluster galaxies (BCGs).  It is believed that the excess UV signal is produced by old and evolved core-He burning stars, and the UV flux strength could be greatly enhanced if the progenitor stars have high value of He abundance.  In this work, we propose that sedimentation process can greatly enhance the He abundance in BCGs.  Our model predicts that the UV flux strength is stronger in more massive, low-redshift, and dynamically relaxed BCGs.  These predictions are testable with the current generation of GALEX+SDSS observations. 
\end{abstract}

\keywords{galaxies: elliptical and lenticular, cD --- galaxies: clusters: general --- ultraviolet: galaxies --- diffusion}
\section{Introduction}
\label{sec:intro}

Brightest Cluster Galaxies (BCGs) are among the most massive galaxies in the Universe, forming and evolving at centers of galaxy clusters through galaxy mergers and accretion.  The baryonic component of present-day BCGs is composed of mainly old stellar populations \citep[e.g.][]{DeLucia07, Collins09} with a fraction of younger stellar populations formed out of cooling gas in cluster cores \citep[e.g.,][]{Peterson06}.  Properties of BCGs are therefore dependent on the mass accretion histories and the conversion of hot gas in clusters into stars.  Detailed studies of the BCGs properties have the potential to shed new insights into the physics of the most massive galaxies in the Universe.

The UV upturn is one of the mysterious phenomena observed in massive elliptical galaxies.  Since the first discovery \citep{code_welch79}, it has been known for several decades that some giant elliptical galaxies exhibit strong unexpected bump in the Ultraviolet (UV) part of the spectrum \citep[see][for a review]{OConnell99}.  Recent GALEX observations indicate that the UV flux excess is most significant in BCGs \citep{ree_etal07}. The observed UV excess, also known as the UV upturn, is spatially smooth and extended \citep[e.g.,][]{lee_etal05a,ree_etal07}.  The spatial distribution and the spectral shape of the UV upturn phenomena cannot be explained by recent star formation \citep{brown_etal97}.  

One of the leading models of the UV upturn suggests that the phenomena can be explained most naturally by old evolved stars, known as extreme horizontal-branch (EHB) stars.  EHB stars are hot, core helium-burning stars with extremely thin hydrogen envelopes ($M_{\rm env} \leq 0.05 M_\sun$), which lost their envelopes via massive winds on the red giant branch in the single star evolution model \citep[e.g.,][]{Dorman93} \footnote{It has also been suggested that EHB stars can be produced in binary star systems, where the envelope of the core He-burning star can be tidally stripped by a companion star \citep{Han07}.}.  

High value of He abundance can make the formation of hot EHB stars more effective \citep{Dorman95,Yi97}. This is because that HB stars with a higher helium abundance evolve faster on main sequence and red giant branch, can burn more hydrogen during the core helium-burning phase, and therefore can become UV-bright more easily.  In particular, a very large He abundance ($Y \gtrsim 0.4$) can greatly boost the UV upturn strength \citep{Dorman95}.

In this work, we point out that helium sedimentation process can greatly increase the helium abundance in BCGs, making them conducive to the formation of UV-bright EHB stars.  It has long been suggested that helium sedimentation occurs in the intracluster medium (ICM) of galaxy clusters \citep{Abramopoulos1981On_the_equilibr,Gilfanov1984Intracluster,Qin2000BARYON-DISTRIBU,Chuzhoy2003Gravitational,Chuzhoy2004Element,Ettori2006Effects,Peng_Nagai09}. Under the influence of gravity, the heavier helium nuclei in the H-He dominated ICM accumulate at the center of massive galaxy clusters.  If the UV flux is produced mainly by EHB stars, the He sedimentation process can boost the UV flux produced by these stars.   Our sedimentation scenario predicts that the UV upturn phenomena should be most pronounced in high-mass (or $T_X$), low-redshift, and dynamically relaxed systems.  We show that these predictions are testable with the current generation of GALEX+SDSS observations.

\section{He sedimentation in Galaxy Clusters and Elliptical Galaxies}
\label{sec:sediment}

\subsection{Diffusion Equations} 
\label{sec:diffusioneqn} 

We calculate the evolution of the He sedimentation in galaxy clusters and elliptical galaxies by solving the diffusion equations for a fully ionized hydrogen and helium plasma.  Detailed descriptions of these calculations are described in \citep{Peng_Nagai09}.  Here we provide a brief overview of the calculations. 

Each species $s$ obeys an equation of continuity and momentum conservation,
\begin{eqnarray}
  \frac{\partial n_s}{\partial t}+ \frac{1}{r^2}\frac{\partial (r^2n_s u_s)}{\partial r} &=&0~
  \label{continuity.e}\ ,\\
  \frac{\partial P_s}{\partial r} + n_s A_s \mb g - n_s Z_s eE &=& \sum_t K_{st} (w_t - w_s) \ .
  \label{burgers.e}
\end{eqnarray}
Here the species $s$ has mass $A_s \mb$, charge $Z_s e$, density $n_s$, partial pressure $P_s$, and velocity $u_s$, where $\mb$ is the proton mass.  The effective resistance coefficients is defined as $K_{st} \equiv f_B^{-1} K_{st}^{B=0}$, where $f_B$  is a magnetic or turbulent suppression factor and $K_{st}^{B=0}  \propto T^{-3/2}$ is the resistance coefficient of the un-magnetized H-He plasma \citep{chapman.cowling}.  In the following calculations, we assumed that $f_B =1$ to obtain the maximum He sedimentation effect. Equation~(\ref{burgers.e}) describes forces acting on a specie $s$, and it is the balance of these forces that ultimately determines the rate of sedimentation.  For a sinking He nucleus, the gravitational force ($g$) is counteracted by three types of forces provided by the induced electric field ($E$), the pressure gradient ($dP_s/dr$) of helium, and the drag force due to collisions with surrounding particles.  Note that the sedimentation destroys hydrostatic equilibrium since redistribution of particles introduces a temporal change in the total gas pressure.  However, hydrostatic equilibrium can be restored quickly.  This equilibrium restoring acquires a net inflow with a mean velocity $u = \sum_s n_s A_s u_s/\sum_s n_s A_s$,
\beq \frac{du}{dt} = - \frac{1}{\rho\gas} \frac{\partial
P\gas}{\partial r} - g \ ,
\label{du.e}
\eeq
where $P\gas = \sum_s n_s k_{\mathrm B} T$ is the total gas pressure for ideal gas, and $\rho\gas = \sum_s n_s A_s m_u$ is the gas density. 

The diffusion velocity between species $s$ and the fluid element is $w_s = u_s - u$, which satisfies  mass and charge conservation,
\begin{eqnarray}
  \label{mc.e}
  \sum_s A_s n_s w_s &=& 0 \ , \\
  \label{cc.e}
  \sum_s Z_s n_s w_s &=& 0 \ .
\end{eqnarray}
Note that the summations include both ions and electrons. To satisfy  these conservation laws, for each sinking helium nuclei, there are roughly four protons and two electrons that float up.  Equations~(\ref{continuity.e})-(\ref{cc.e}) describe the process of particle diffusion in a multi-species plasma.  

\subsection{Cluster and Galaxy Models} 
\label{sec:model} 

We set up cluster and galaxy models and initial conditions as follows. Initially, we assume that the gas consists of a primordial H and He plasma uniformly throughout clusters (or galaxies) ($X = 0.75$ and $Y = 0.25$).  We ignore the contribution of elements heavier than He.  We set up the initial gas distribution by assuming hydrostatic equilibrium of the plasma medium in the potential well of clusters dominated by dark matter, following the Navarro-Frenk-White (NFW) density profile for the total mass distribution \citep{Navarro1997A_Universal_Den}. We adopt the concentration parameter $c_{500} \equiv r_{500}/r_s = 4$ for clusters \citep{Vikhlinin2006Chandra_Sample}(V06) and $10$ for elliptical galaxies \citep{Wechsler2002}, where $r_s$ is the scale radius of the NFW density profile and $r_{500}$ is a radius of a spherical region within which the mean enclosed mass density is $500$ times the critical density of the universe. 

For the temperature profile, we consider the observed temperature profile obtained with deep Chandra observations of nearby relaxed clusters (V06). The observed temperature peaks around $r/ r_{500} \simeq 0.2$, and decreases at both inner and outer radii \citep[see Figure~1 in][]{Peng_Nagai09}. The observed temperature drop in the inner radii leads to a significant suppression of He sedimentation, compared to the results based on the isothermal model \citep{Chuzhoy2004Element}. For elliptical galaxies, due to their lower X-ray luminosity, the temperature observations are more difficult and usually limited to core regions. We assume that they have the same universal temperature profiles of clusters. Most of the observed bright elliptical galaxies do show V06-like temperature profiles \citep{2008Diehl, 2006Fukazawa}. The gas density is derived from solving the hydrostatic equilibrium equation. The central density is normalized by requiring the enclosed gas mass fraction to be $f_{\rm gas}(<r_{500}) = 0.15$ for clusters (V06) and $0.015$ for elliptical galaxies.\footnote{The baryon components of elliptical galaxies are dominated by stars.}  Throughout this work, we use cosmological parameters: $\Omega_{\rm M} = 0.3$, $\Omega_\Lambda = 0.7$, $\Omega_{\rm b} = 0.0462$, and $h = 0.7$ \citep{2008Komatsu_Five_Year}. 

\subsection{Sedimentation Velocity and Scaling Relations}
\label{sec:analytic}

To develop physical insights into the process of He sedimentation, it is useful to consider a drift velocity of a trace He particle in a background of hydrogen, i.e, $n\hy \gg n\he$ 
and $w\hy = 0$.  In this limiting case, equations~(\ref{burgers.e}) for the two species are decoupled.  The right-hand side of the equations for H vanishes, thereby fixing an electric field, $eE = 0.5\,\mb g$.  Substituting $E$ into the equation of motion for He, we obtain the sedimentation velocity of He nuclei as $w\he = 3\,\mb g n\he /K_{\rm{pHe}}$, which gives
\beq
w_{\rm He}  \simeq 80\usp{\rm km\,s^{-1}} \left(\frac{T}{10\usp{\rm keV}}\right)^{3/2} 
\left(\frac{g}{10^{-7.5}\usp{\rm cm\,s^{-2}}}\right) \left(\frac{n}{10^{-3}{\rm cm^{-3}}}\right)^{-1}
\label{wsed}
\eeq
where the induced electric field counteracts gravity and suppresses the sedimentation speed by $25\%$.  At a fixed density, the sedimentation velocity is generally larger for higher temperature ($T$) and gravity ($g$).  Note that the pressure gradient ($dP/dr$) term in equation~(\ref{burgers.e}) further suppresses the sedimentation process.  

Substituting Eq.~(\ref{wsed}) into Eq.~(\ref{continuity.e}) and integrating over time, we obtain $Y =  Y_0 e^{4\Delta t/3\tau}$, where $Y_0$ is the primordial He mass fraction and $\tau \simeq 10 \, {\rm } {\rm Gyr}\, (f\gas/0.1) (T/10\, {\rm keV})^{-3/2} (1 + 1.5\, d\ln T/d\ln r)^{-1}$ is the sedimentation timescale  \citep{Chuzhoy2004Element}.  Note that the gas temperature gradient term, $d\ln T/d\ln r$, takes into account of the non-isothermal temperature profile.  For $t << \tau$, the He enhancement factor, $\Delta Y \equiv Y - Y_0$, has the following scaling $\Delta Y \propto \Delta t/\tau \propto \Delta t f\gas^{-1} T^{3/2}$.  This indicates that the He sedimentation process is efficient for systems with higher $T$ and lower $f_{\rm gas}$.

\subsection{He Abundance Enhancement in BCGs}
\label{sec:results}

\begin{figure}[t]
  \centering
  \includegraphics[clip=true, trim=0.0cm 0.0cm 0.0cm 0.0cm, width=4.0in]{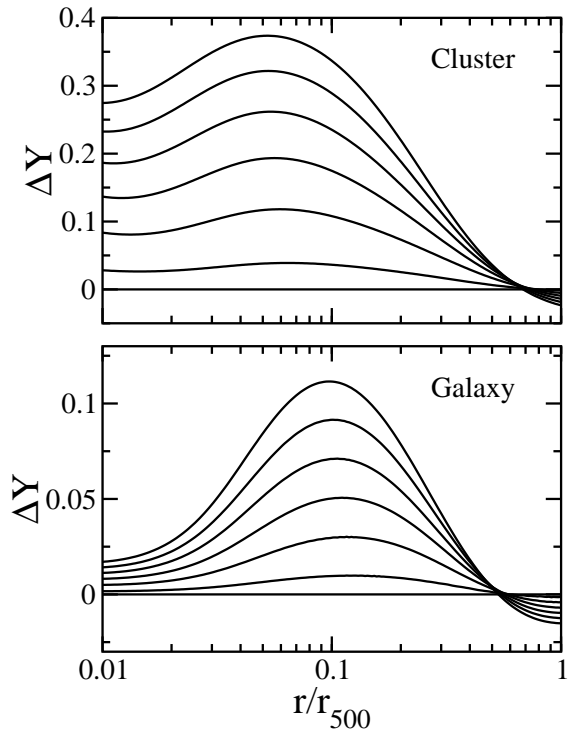}
   \caption{The spatial distribution of He enhancement factor in a $T_X = 10$ keV cluster (\emph{top panel}) and in a $T_X = 1$ keV  elliptical galaxy (\emph{bottom panel}). The set of curves in each panel, starting from the flat line, are for ages of 0, 1, 3, 5, 7, 9 and 11 Gyr, respectively. For galaxy clusters, we assumed $f_{\rm gas}(<r_{500})=0.15$ and $c_{500} = 4$. For elliptical galaxies, we assumed $f_{\rm gas}(<r_{500})=0.015$ and $c_{500} = 10$. }
  \label{Y_r.f}
\end{figure}

In this section, we solve the diffusion process in the H-He plasma numerically.  Figure~\ref{Y_r.f} shows the spatial distribution of the He enhancement factor, $\Delta Y$, for a $T_X=10$~keV galaxy cluster and a $T_X = 1$~keV elliptical galaxy.  The figure shows that the He enhancement could be significant ($\Delta Y \sim 0.3-0.4$) in the core of massive, relaxed clusters.  For clusters, the He enhancement factors peak around $r \simeq 0.05\,r_{500}$ and reaches $0.4\, (0.2)$ for cluster age of $11\, (5)$ Gyrs.   For elliptical galaxies, the He enhancement is also non-negligible  ($\Delta Y \sim 0.1$), because of their low gas mass fraction.  Here, the He enhancement factor peaks around $0.1\, r_{500}$ with $\Delta Y = 0.1\,(0.05)$ after $11\,(5)$ Gyrs.  Because of their higher concentration parameter, the elliptical galaxy has significant suppression of He sedimentation in the galaxy central core region. This is because that the higher pressure gradient causes more efficient outward He diffusion and hence suppresses the process of He sedimentation.

\begin{figure}[t]
  \centering
\includegraphics[clip=true, trim=0.0cm 0.0cm 0.0cm 0.0cm, width=4.0in]{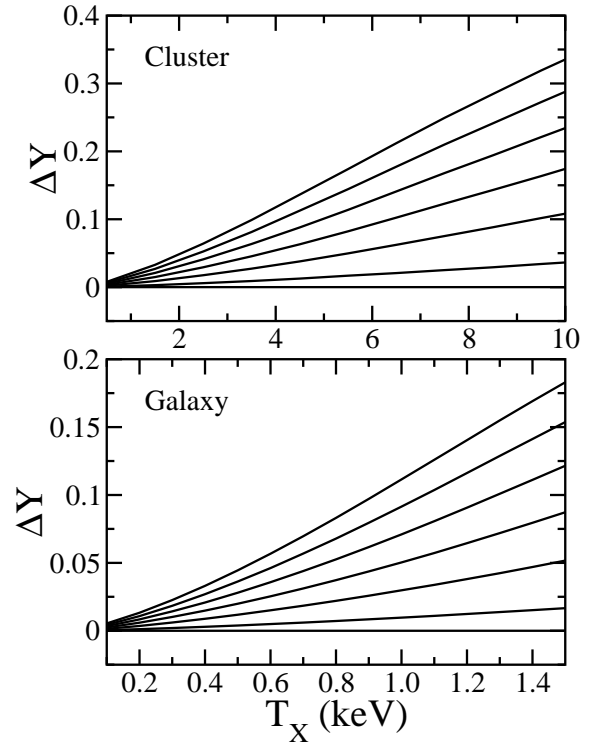}
   \caption{The dependence of the He mass fraction at $r = 0.1\,r_{500}$ on
  X-ray temperature of galaxy clusters (\emph{top panel}) and elliptical galaxies
  (\emph{bottom panel}).  The curves starting from the flat line are for ages of 
  0, 1, 3, 5, 7, 9 and 11 Gyr, respectively.}
  \label{Y_T.f}
\end{figure}

Figure~\ref{Y_T.f} shows the He enhancement at $r = 0.1\,r_{500}$ as a function of X-ray temperature of galaxy clusters (\emph{top panel}) and elliptical galaxies (\emph{bottom panel}).  Note that $0.1\,r_{500}$ roughly corresponds to the effective (half-light) radius of stellar surface brightness for BCGs.\footnote{For BCGs of stellar mass $10^{12}\usp M_\sun$, $0.1\,r_{500} \sim 50$ kpc and the effective radii $\sim 30-100$ kpc \citep{Graham06}.}  Figure~\ref{Y_T.f} demonstrates that the He sedimentation is strongly sensitive to the X-ray temperature.  The He enhancement is also inversely proportional to the gas mass fraction, leaving the He sedimentation in elliptical galaxies finite.  

Since the BCGs are galaxies located at the center of clusters, the He enhancement in BCGs is roughly given by the sum of the effects in clusters and elliptical galaxies: $\Delta Y \sim 0.5(0.25)$ after $11(5)$ Gyrs. Note that this is the optimal He enhancement in BCGs since the He sedimentation process could be suppressed by magnetic field and turbulence. 

\section{Implication for the UV phenomena}

In \S~\ref{sec:results}, we showed that He sedimentation can greatly enhance the He abundance in BCGs and that the effect is very sensitive to the X-ray temperature of host galaxy clusters.   As we discussed in Section~\ref{sec:intro}, high He abundance can make the environment conducive to the formation of the UV-bright EHB stars  \citep{Dorman95}.  Thus, the He sedimentation may provide an attractive mechanism to explain why the UV upturn phenomena is pronounced in BCGs.  

Connecting the idea of the He sedimentation to observations of the UV upturn phenomena is a nontrivial step that requires detailed modeling of star formation and chemical enrichment histories as well as the full stellar population synthesis calculation that take into account the EHB formation and the He sedimentation.  The full self-consistent modeling is thus beyond the scope of this work.  
However, it would be useful to check if the He sedimentation can have prominent effect on boosting  the UV flux in BCGs, more specifically, the observed Far-UV (FUV)-to-V flux ratio, also known as the UV flux strength. The FUV flux is contributed mainly by stars in the HB/post-HB phases, while the V flux is contributed mainly by main-sequence stars.  The FUV-to-V flux ratio is therefore given by $L_{FUV}/L_V = \dot{n}(t) \epsilon_{FUV}$, where the total FUV flux is governed by the rate of stars entering the HB phases (or leaving off the main sequence) ($\dot{n}(t)$)\footnote{The number of stars leaving main sequence per unit time per unit $L_V$.} and the energy output in FUV band over the lifetime of HB/post-HB evolutionary track ($\epsilon_{FUV}$).  

High He abundance can greatly enhance the UV energy outputs $\epsilon_{FUV}$ \citep{Dorman95}.  An increase of He abundance yields a smaller main-sequence turnoff mass at a given age, and therefore a smaller envelope mass.  It was demonstrated that sufficiently He-enhanced younger stellar population can still have smaller Main Sequence (MS) turnoff mass than an older population with a normal He abundance \citep[see e.g.,][Table 4]{Dorman95}. A larger He abundance can also cause stars with an given envelope mass to burn more of their hydrogen envelope during their He core burning phase.  Helium-rich stars can therefore produce UV flux for a larger range of initial envelope masses.  For example, increasing $Y$ from $0.27$ to $0.47$ roughly increases the total FUV energy output by a factor of 4 for a uniform distribution of envelope mass \citep{Dorman95}.  $\dot{n}(t)$, on the other hand, is fairly insensitive to the compositions. For the range of He abundance considered in this work, the enhancement factor of the UV flux strength due to high He abundance is $\leq 4$.  

Assuming that most of the stars are formed in a star burst at an early redshift (denoted as 'org'), and the rest is formed continuously with a constant star formation rate out of He-enriched gas (denoted as 'con'), we can express the total FUV-to-V flux ratio as
\begin{eqnarray}
\label{enhance_1.e}
\frac{L_{FUV}}{L_V} & \simeq  & \left(\frac{L_{FUV}}{L_V}\right)_{\rm org} + f_{\rm con} \left(\frac{L_{FUV}}{L_V} \right)_{\rm con}  \\           
           & = & \left(\frac{L_{FUV}}{L_V}\right)_{\rm org}(1 + \alpha f_{\rm con}) .
           \label{enhance_2.e}
\end{eqnarray}
where $f_{\rm con}$ is the ratio of the continuous forming stellar mass to the total stellar mass, and $\alpha$ is the enhancement factor of UV flux strength due to high He abundance. In deriving Eq. (\ref{enhance_1.e}), we assumed that the visual flux is not significantly affected by the He abundance. This is reasonable since the young He-enhanced stellar population has similar MS turnoff mass as the old first-generation stellar population. 

While it is widely believed that AGN feedback may significantly suppress star formation in most massive elliptical galaxies \citep{Pipino09a, Schawinski06}, observations from the ultraviolet fluxes \citep{Hicks05, Hicks09, Pipino09b, Wang09}, the infrared fluxes \citep{Egami06, ODea08}, optical photometry \citep{Bildfell08, Rafferty08} and optical line emission \citep{Crawford99, Edwards07} of the BCGs suggest that recent star formation is still taking place in some of these objects.  In particular, about one third of the X-ray selected BCGs associated with cool-core clusters exhibit recent star formation.  This indicates that AGN feedback may not completely suppress star formation in BCGs at the center of cool-core clusters.  Here we assume that 10-20\% of stars may form from cooling gas at cluster centers.\footnote{The value is reasonable for BCGs with $30\usp M_\sun \,{\rm yr}^{-1}$ of star formation rate \citep{Peterson06,ODea08} and a typical stellar mass $10^{12}\usp M_\sun$.}  Using $\alpha = 4$ and $f_{\rm con} = 0.1$ for the present-day BCGs, we expect that the UV flux would be enhanced by $40\%$ due to the He sedimentation effect (from eq.~[\ref{enhance_2.e}]). The $FUV - V$ color, therefore, is $\sim 0.4$ mag bluer.   

Therefore, one of the predictions of our model is that the effect of He sedimentation, and hence the UV flux, is the largest for the BCGs in cool-core clusters than in non-BCG elliptical galaxies.  Systematic studies of the UV flux strength in BCG and non-BCG elliptical galaxies can test our models. 

On smaller scales, recent observations suggest that some globular clusters (GCs) in our Galaxy, such as $\omega$ Cen, requires He-enhanced population ($Y \simeq 0.4$) to explain both the blue main-sequence population as well as the hot horizontal branch stars \citep{Lee_etal05b, Piotto05}.  However, we point out that the global He sedimentation process is likely unimportant for GCs.  It is because that GCs pass through the Galactic disk on a timescale of $\sim 0.1$ Gyr and each cluster-disk interaction would wash out the accumulated gas in the GCs via ram pressure stripping. Since the interaction interval is much less than the He sedimentation timescale ($\sim 1$ Gyr for a $10^6 M_\odot$ GC), there is not sufficient time for He sedimentation to build up significantly high He abundance.

\section{Conclusions and Discussions}
\label{sec:conclusion}

In this work, we propose that the He sedimentation can significantly enhance the UV signal observed in BCGs.  We show that the He sedimentation process can produce significant amount of He abundance ($\Delta Y>0.25$) in central regions of hot, massive galaxy clusters. Stellar populations in BCGs may contain UV bright EHB stars when they are old.  The model predicts that the UV upturn phenomena increases with X-ray temperature or mass of galaxy clusters.  

Our sedimentation model makes several unique predications that are testable with current GALEX+SDSS observations.  

\newcounter{bean}
\begin{list}
{\arabic{bean}.}{ \usecounter{bean} \setlength{\parsep}{+0.03in}
\setlength{\leftmargin}{+0.15in} \setlength{\rightmargin}{+0.15in}}

\item{}
The model predicts a strong correlation between the UV upturn strength of BCGs and the X-ray temperature of host galaxy clusters. 

\item{} 
The model also predicts that the UV upturn phenomena should be more pronounced in BCGs, compared to non-BCG elliptical galaxies.  

\item{}
Some correlations are expected between the UV upturn strength of BCGs and the dynamical state of the BCGs and their host galaxy clusters.  Major mergers, for example, can destroy He-rich cluster cores, producing some scatter in the UV upturn among BCGs.  In this case, we expect a correlation between the size of UV upturn and dynamical state of clusters (e.g., measured using X-ray morphology or multiple X-ray peaks).  This is another expected observational signature of our model.

\end{list}

Analyses of a large sample of galaxies from the GALEX+SDSS along with average temperatures of their host X-ray clusters can provide important test of this model and may shed new light on the origin of the UV upturn phenomena as well as the physics of the formation and evolution of massive galaxies in the universe.

\acknowledgements

We thank Sarbani Basu, Thibaut Decressin, Pierre Demarque, Philipp Podsiadlowski, Kevin Schawinski, and Sukyoung Yi for useful discussions. We thank the anonymous referee for helpful comments.  This work is supported by Sherman Fairchild Foundation.

\clearpage


\end{document}